\documentstyle[11pt,aaspp4]{article}

\newcommand\kms  {{km~s$^{-1}${}}}
\newcommand\hmpc {{$h^{-1}\,$~Mpc{}}}
\newcommand\rt   {{\tilde R}}                      
\newcommand\mrt  {{$\tilde R$ }}
\newcommand{\abf} {{Ab$_{135}$}}
\newcommand{\abs} {{Ab$_{137}$}}

\begin{document}

\title{Kinematics and Mass Profile of AWM 7\footnote{
Observations reported in this paper were obtained at the
Multiple Mirror Telescope Observatory, a facility operated jointly
by the University of Arizona and the Smithsonian Institution; at the Whipple
Observatory, a facility operated jointly by the Smithsonian Astrophysical
Observatory and Harvard University, and at the WIYN Observatory, a joint
facility of the University of Wisconsin-Madison, Indiana University,
Yale University, and the National Optical Astronomy Observatories.}}

\author{Daniel M. Koranyi and Margaret J. Geller}
\affil{Harvard-Smithsonian Center for Astrophysics}
\authoraddr{60 Garden St., Cambridge, MA 02138}
\authoremail{dkoranyi@cfa.harvard.edu, mgeller@cfa.harvard.edu}

\begin{abstract}

We have measured 492 redshifts (311 new) in the
direction of the poor cluster AWM~7 and have identified 179 cluster
members (73 new).  We use two independent methods to derive a
self-consistent mass profile, under the
assumptions that the absorption-line galaxies are virialized and that
they trace an underlying Navarro, Frenk \& White (1997) dark matter profile:
(1) we fit such an NFW
profile to the radial distribution of galaxy positions and to the velocity
dispersion profile; (2) we apply the virial mass estimator to the cluster.
With these assumptions, the two independent mass estimates agree to
$\sim$15\% within 1.7 \hmpc, the radial extent of our data; we find an
enclosed mass $\sim$$(3\pm0.5)\times 10^{14}\,h^{-1}\,$M$_\odot$.
The largest potential source of systematic error is the inclusion
of young emission-line galaxies in the mass estimate.

We investigate the behavior of the surface term correction to the
virial mass estimator under several assumptions about the velocity
anisotropy profile, still within the context of the NFW model, and
remark on the sensitivity of derived mass profiles to outliers.  We
find that one must have data out to a large radius in order to
determine the mass robustly, and that the surface term correction is
unreliable at small radii.

\end{abstract}


\section{Introduction}

The properties of galaxy clusters play an important role in constraining
cosmological parameters and models of large-scale structure formation.
Because clusters are the most massive collapsed aggregates in the universe,
the cluster mass function and its evolution with redshift constrains
the spectrum of density fluctuations (e.g. Eke, Cole \& Frenk 1996, Bahcall
\& Cen 1992).  Accurate cluster masses can also constrain the ratio of
baryonic to total mass, and thus $\Omega_0$ (White et al. 1993).

There are several standard ways of computing cluster masses.  Optical redshift
surveys yield a mass estimate through the application of the virial
theorem, on the assumption that the galaxies trace the overall cluster
mass distribution; this method was first applied to Coma by Zwicky (1933).
More sophisticated variants of this method include a surface term to 
account for the unsampled portion of the cluster (The \& White 1986), and 
use the Jeans equation to account for the effects of orbital anisotropy
(e.g. Binney \& Tremaine 1987).  For clusters with many hundreds of 
redshifts, the distribution of infalling galaxies in redshift space
measures the gravitational potential of the dark matter
halo (Diaferio \& Geller 1997); this method has only been successfully
applied to Coma so far (Geller, Diaferio \& Kurtz 1999).  Observations of hot
intracluster gas in the X-ray provide information about the gravitational
potential; the added assumption of hydrostatic equilibrium yields a
mass (Bahcall \& Sarazin 1977).  Gravitational lensing of background
galaxies can also be used to estimate cluster masses (Webster 1985,
Bartelmann et al. 1996 and references therein).  Comparing cluster masses
derived by these independent methods can provide insight
into the internal structure of clusters (Loeb \& Mao 1994) and provides
a consistency check on the mass derived by any one method.  

While the distribution of total cluster masses constrains global cosmological
parameters, knowledge of the density (or equivalently, mass) profiles of 
individual
clusters constrains models of cluster formation.  $N$-body simulations of
hierarchical clustering by Navarro, Frenk \& White (1995,1996,1997, hereafter
NFW) suggest
that a universal two-parameter density model can adequately describe
clusters over a broad range of masses.  Until recently, few clusters had
enough measured redshifts to constrain the mass profile; the
advent of multi-fiber spectrographs has now made it possible to measure several
hundred redshifts in a cluster within a reasonable time.

Our goal is to measure the mass profile of the cluster AWM~7, selected
by Albert et al. (1977) on the basis of the cD galaxy at its center.
Our sample consists of 179 cluster galaxies with measured redshifts;
this work expands the sample of Koranyi et al. (1998) by 70\%,
extending the surveyed region to a projected radius of 1.7 \hmpc, and reaching
fainter magnitudes in the core.  We confirm the cold core seen in
Koranyi et al. (1998), and fit a NFW mass profile from the projected
distribution of radii for comparison with the virial mass profile;
we consider also the effect of orbital anisotropies (still assuming
the NFW profile) on the mass estimates.
The two profiles agree remarkably well within our restricted context;
the assumption of the NFW model produces self-consistent mass estimates
using these two methods.  Relaxing the assumption that the galaxies 
trace an underlying NFW dark matter distribution would result in a
larger range of admissible masses (cf. The \& White 1986).

In \S 2 and \S 3, we describe the observations and data reduction.  In
\S 4 we describe the data, and in \S5 we justify the choice of the NFW
model, compute the resultant mass profiles,
and comment on the reliability of the surface term
correction.  We defer discussion of some details until \S 6,
and conclude in \S 7.  We use $H_0 = 100\,h$~km~s$^{-1}$~Mpc$^{-1}$ 
throughout.

\section{Observations}

We observed a total of 593 distinct targets in the direction of AWM~7 from
1995 to 1999.  We measured redshifts for 492 objects; 179
are cluster galaxies (with redshifts in the range 2500--7500 \kms),
making AWM~7 one of the best-sampled clusters in
the sky.  Our sample extends over 1.5 degrees in radius from the cluster
center, and reaches magnitudes as faint as $R = 18$ in the core.
We collected data from three different telescopes,
with different sky coverage and different magnitude limits;
we describe the details of each in turn.  In each case we chose the spectral
coverage so that both the Ca H-K break at $\lambda\lambda3933,3968$ \AA\ and 
H$\alpha$ $\lambda6562$ \AA\ would fall on the detector at the cluster
redshift.  These spectral features are important for determining
redshifts by cross-correlation with template spectra.  We list the 
redshifts for all measured cluster galaxies in Table~\ref{tab:cz}.

\subsection{FAST Observations}

We used the FAST Spectrograph (Fabricant et al. 1998) mounted on the
1.5-m Tillinghast telescope at the F.L. Whipple Observatory on Mt. Hopkins
to observe 302 targets over $\sim$10 square degrees of sky centered on 
NGC~1129.  We selected the targets from the Digitized
Palomar Observatory Sky Survey scans.  We used a 3 arcsecond slit and a
300 line~mm$^{-1}$ grating, resulting in a resolution of $\sim$6 \AA\
and spectral coverage of 3600--7600 \AA.  FAST can reliably measure redshifts
for galaxies down to $R \sim 15.5$, and can measure redshifts for galaxies
as faint as $R\sim16$ if the central surface brightness is sufficiently high.
159 FAST targets are new to this paper; 51 of these are cluster members.

\subsection{MMT Observations}

We observed 45 targets with the Blue Channel spectrograph of the MMT on
the nights of 1996 Dec. 5 and 6, using a 2 arcsecond slit and a 300
line~mm$^{-1}$ grating, resulting in a resolution of $\sim$10.6 \AA\
and spectral coverage of 3000--8000 \AA.  These targets are in the central
$75' \times 75'$ $(1.15 \times 1.15 h^{-1}$~Mpc) of the cluster, and 
are complete to $R=16.5$ in that region.  These data have been previously
published in Koranyi et al. (1998).

\subsection{Hydra Observations}

We observed 250 targets (under the queue observing scheme) with the Hydra
multi-fiber bench spectrograph mounted on the WIYN telescope at Kitt Peak
on the nights of 1998 Dec. 20 and 21; these data are new to this paper.
There were 4 separate fiber
configurations; 60 objects were observed in more than one configuration.
There were either four or five 25-minute exposures for each configuration.
We used 3 arcsecond fibers, and a 400 line~mm$^{-1}$ grating, resulting in
a resolution of $\sim$7 \AA\ and spectral coverage of 3800--7000 \AA.  The 
four configurations were observed under 
strongly varying conditions, resulting in dramatic differences in throughput.
Two of the configurations suffered from wind shake, and a third from cirrus.
The remaining configuration was observed under stable, clear skies with good
seeing and resulted in a redshift for almost every target.

In total, we were able to measure redshifts for 154 of the 250
targets (152 new), sometimes by combining
spectra of the same object observed under different configurations.  Only
22 of these are cluster members.  The Hydra targets are all within the
central $1.4^\circ \times 1.4^\circ$ of the cluster.

The magnitude completeness of the Hydra sample is complicated.  Not only were
we unable to measure a redshift for every target, but the original target 
list itself did not include every galaxy (up to some limit) in a
magnitude-ordered list because of physical constraints in allocating 
fibers to objects.  Moreover, the initial magnitude ranking was not by
total galaxy luminosity, but rather by luminosity within the central 3
arcseconds subtended by the fibers.
We will discuss the photometry and luminosity function
of AWM~7 in a later paper; here, we are interested primarily
in the kinematics of the cluster and the incompleteness is not a limitation
on the analysis.

\section{Data Reduction}

We reduced all spectra using the {\bf xcsao} task of the RVSAO package
in IRAF (Kurtz \& Mink 1998).  {\bf xcsao} measures redshifts by
cross-correlating the observed spectra against a suite of template
spectra.  The quality of the match is measured by the parameter $R$, an
estimate of the ratio of the highest peak to the average noise peak in
the correlation function.  The template spectra are themselves
empirically derived from an archive of FAST observations, and are thus
strictly speaking directly applicable only to other FAST spectra.  For
the FAST velocities, the instrumental velocity offset with respect to
the night sky lines is less than 10 km~s$^{-1}$ and is
indistinguishable from zero (Kurtz \& Mink 1998).  We have found that
the FAST template spectra are reliable also for the MMT and Hydra
spectra; for those objects observed with several spectrographs, the
redshifts derived from the separate spectra all agree within the
errors.

Many of the Hydra spectra have very low signal-to-noise ratios.  We assess
the robustness of the redshifts as follows: for each object, we construct
all $2^N-1$ possible (non-zero) combinations of spectra
from the $N$ individual exposures.
$N$ is 4 or 5 for most objects, and 8, 9 or 10 for objects observed in
two configurations.  For the few objects observed in 3 configurations, with
$N$ as large as 13, we examine only a subset of the 8191 possible
combinations.  We measure the redshift independently from each of these
combined spectra.  We accept a redshift for a target only if the following
criteria all hold:
(1) The $R$-value increases with the number of individual spectra contained
in the combined spectrum; (2) the velocity of the combined spectra stabilizes
as the number of constituent individual spectra increases; (3) the best
combined spectrum has an $R$-value $>$3.  

The comparatively narrow spectral range of the Hydra observations
means that H$\alpha$ is shifted off the edge of the detector for
$cz \gtrsim 20\,000$~km~s$^{-1}$.  Cross-correlating emission line galaxies
at these redshifts with the standard emission-line template results in 
a spurious redshift because the strongest remaining line in the spectrum
is usually fit to H$\alpha$.  We solved this problem by constructing a new
template that differs from the standard emission-line template only by the
absence of the H$\alpha$-nitrogen complex and the
$\lambda\lambda$~6717,6731 \AA\ sulfur lines.  

We add an extra term of 65~km~s$^{-1}$ in quadrature to the errors of velocities
determined from emission-line templates because the line emitting region may
be offset in velocity from the systemic velocity of the galaxy (Kurtz \& Mink
1998).


Emission-line galaxies preferentially yield redshifts
because an emission-line galaxy can be robustly
matched to an emission template even when the signal-to-noise in the continuum
is too low for a believable correlation with an absorption template.  Thus,
the completeness limit for emission-line galaxies is effectively fainter than
for galaxies with no emission.  The Hydra configuration observed through
cirrus yielded few redshifts, resulting in a brighter effective magnitude
limit; this effect is minor because there
is considerable overlap on the sky among the four configurations.

\subsection{Spectral Classification}

We divide our spectra into two broad classes, emission (Em) and absorption (Ab),
based on the presence or absence of H$\alpha$ in the spectrum.
H$\alpha$ is ``present'' if its equivalent width exceeds 5\AA.  The robustness
of this criterion is clearly a function of the signal-to-noise ratio of 
the spectrum; galaxies with low signal or with intrinsically weak H$\alpha$
emission may be misclassified as absorption-line galaxies if the H$\alpha$
peak does not rise appreciably above the noise.  The misclassification is
all in one direction: although some emission line galaxies may be misclassified
as absorption, no absorption-line galaxy will be mistakenly classified as
having emission.

\section{The Data}

\begin{figure}[p]
\centerline{\epsfxsize=6.0in%
\epsffile{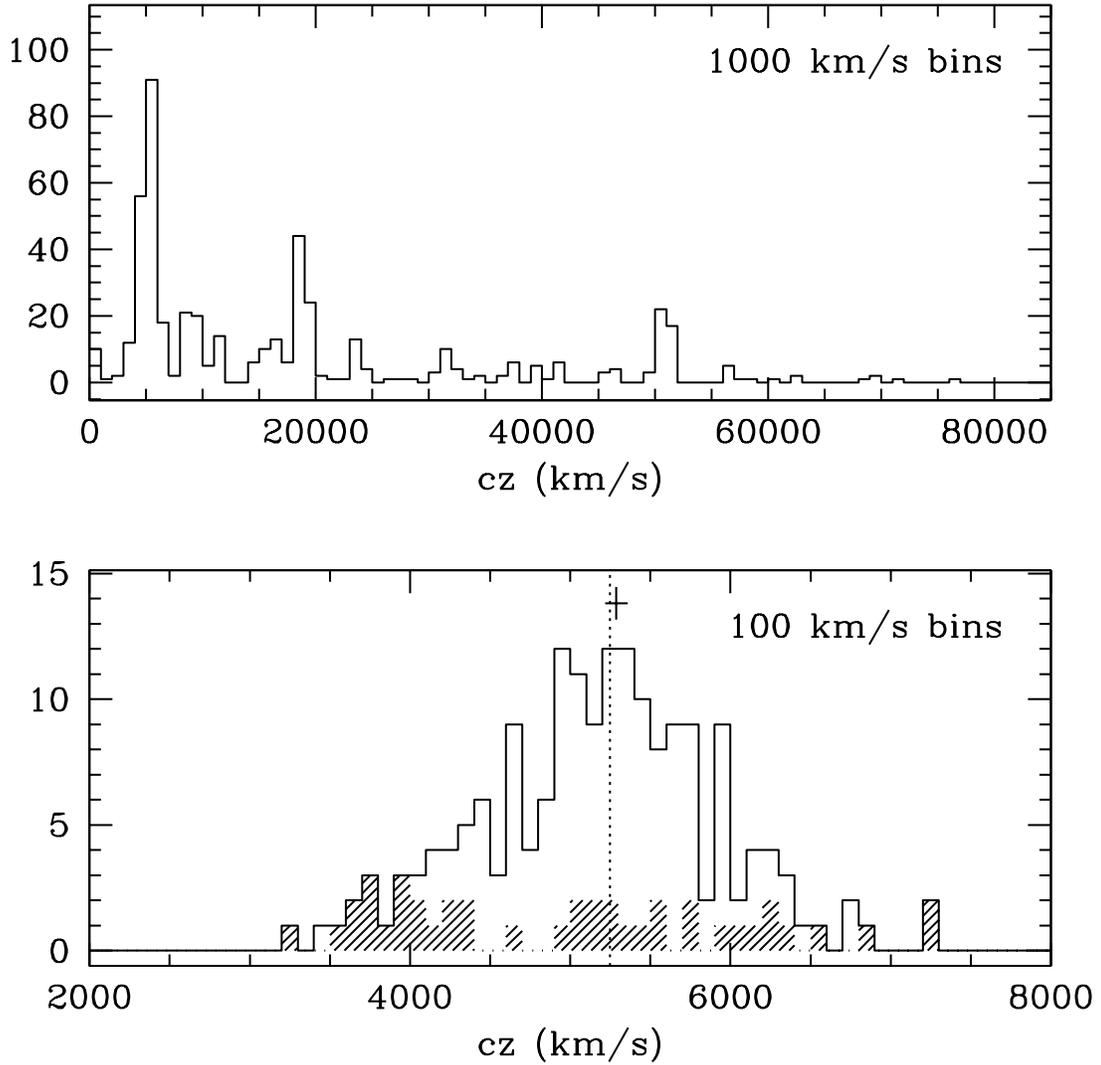}}
\caption{
 Top: Velocity histogram of all observed targets.
 Bottom: Velocity histogram of cluster members.  The solid histogram
  denotes the entire sample (both Em and Ab galaxies); the shaded histogram
  denotes Em galaxies only.  The dotted vertical rule indicates the mean
  redshift of the Ab galaxies; the solid vertical dash over the histogram
  indicates the redshift (and error) of the central cD galaxy.
 }
\label{fig:velhist}
\end{figure}

Figure~\ref{fig:velhist} (upper panel)
shows the velocity distribution of all the observed
galaxies.  The cluster AWM~7, defined as in Koranyi
et al. (1998) to consist of galaxies with redshifts in the range
2500--7500 \kms, is cleanly and unambiguously distinguished
from the background.  The lower panel shows the velocity
histogram of the cluster galaxies only;
the solid histogram indicates the distribution of the entire sample, and
the shaded histogram indicates  the Em galaxies alone.  
Our cluster sample consists of 179 galaxies;
we classify 42 as Em and 137 as Ab.  A Kolmogorov-Smirnov test indicates 
that the probability of the Em and Ab samples being so discrepant if 
drawn from the same distribution is small $(P_{D > D_{\rm obs}} = 0.0004)$;
we conclude that the underlying distributions are different at
the $\sim$4$\sigma$ confidence level.

We compute the mean velocity and dispersion for the different galaxy
populations following the prescription of Danese, De Zotti, \& di Tullio
(1980); the results are in Table~\ref{tab:velhist}.  The Em galaxies have a much
larger dispersion ($\sim$1100 \kms\ vs. $\sim$600 \kms ), and are
systematically offset in redshift by $\sim$400 \kms\ from the Ab
population (4800 \kms\ vs.  5200 \kms).  The extrema of the Em galaxies'
velocity histogram are almost equidistant from the Ab mean velocity, but
near 3800 \kms there is an excess of Em galaxies above the otherwise
almost uniform distribution; this excess decreases the mean redshift.  Neither
the 17 Em galaxies with $3500 < cz < 4500$ \kms nor the 12 with 
$3700 < cz < 4300$ \kms evince any strong clustering on the sky.  Absent
independent distance measurements to these galaxies, it is difficult to
determine whether they are foreground interlopers, or whether the cluster
Em galaxy population is indeed offset in velocity.

The cD galaxy near the cluster center,
NGC~1129, has a redshift of $5288\pm71$~km~s$^{-1}$, which, within the
errors, is at rest with respect to the absorption-line galaxy population.
The peak of the ROSAT PSPC hard X-ray emission map, defined as the
center of the pixel containing the greatest flux, is $6''$ from the nominal
position of NGC~1129.  We therefore adopt its position,
$(\alpha,\delta)_{\rm J2000} = ($02:54:27.50, +41:34:42.50),
as the center of the cluster. 

\begin{figure}[p]
\centerline{\epsfxsize=6.0in%
\epsffile{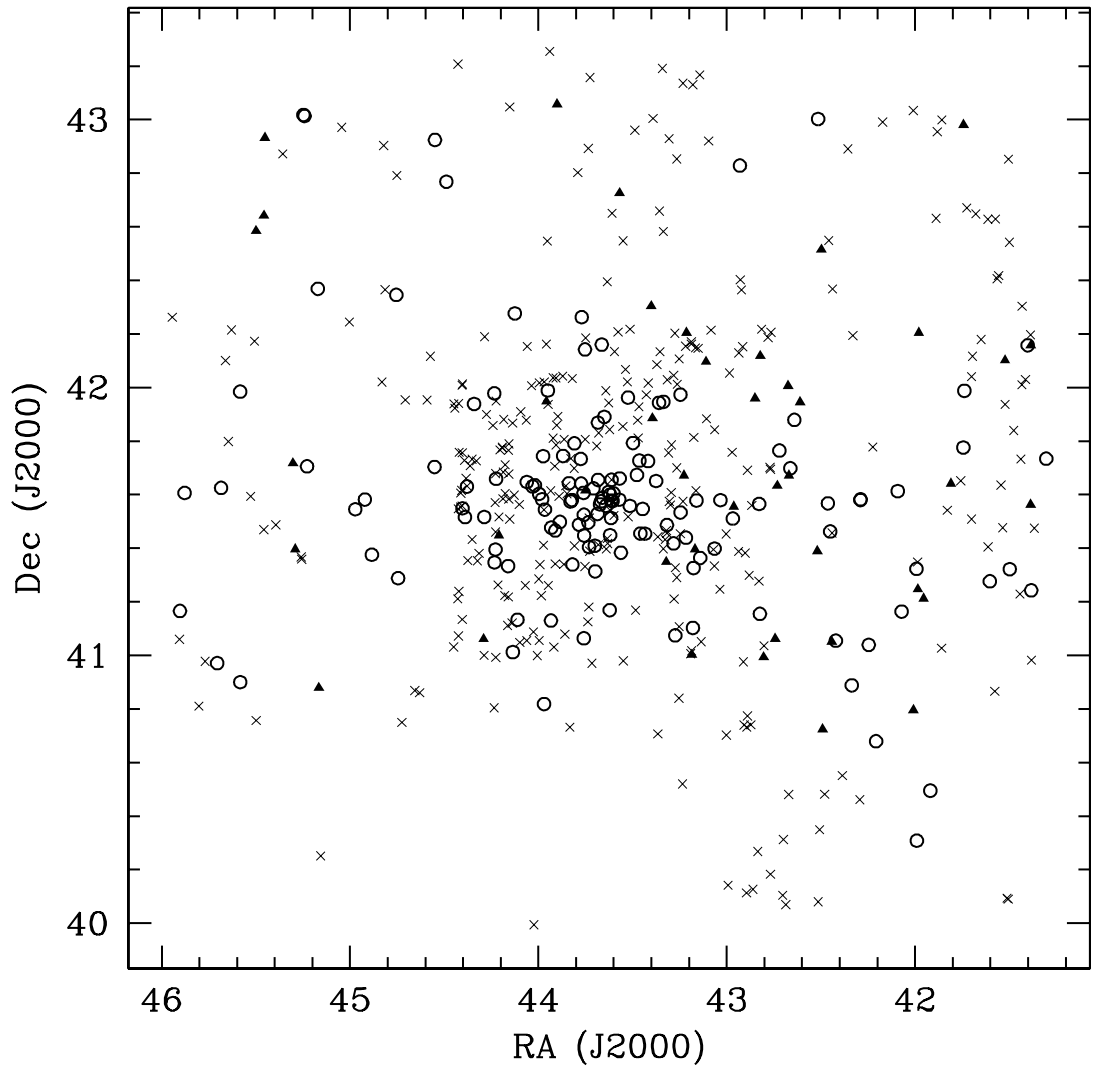}}
\caption{
 Position on the sky of all objects with measured redshifts.  Note the
 denser sampling at the center, in the region covered by the Hydra and MMT
 observations.  Crosses denote foreground and background objects.  Open circles
 denote Ab galaxies in the cluster.  Filled triangles denote Em galaxies in
 the cluster.  Coordinate axes are in decimal degrees.
 }
\label{fig:skypos}
\end{figure}

Figure~\ref{fig:skypos} shows the distribution of galaxies with measured
redshifts on the sky.
Open circles and filled triangles denote Ab and Em galaxies in the cluster,
respectively; crosses denote foreground or background
galaxies.  We do not plot targets for which we were unable to measure a 
redshift.  The enhancement in the surface number density near the center
of the cluster is due to the fainter magnitude limit reached by the MMT
and Hydra observations; most Hydra targets turned out to be background
sources.

The Em galaxies are distributed much more uniformly on the sky than the
Ab galaxies; the projected surface density of the Ab galaxies 
increases strongly toward the center, but the Em galaxies show no such
enhancement.  A two-dimensional Kolmogorov-Smirnov test (Press et al. 1992)
indicates that the probability of drawing two such discrepant samples
from the same underlying distribution is less than $5\times 10^{-5}$.  The
distribution of Ab galaxies is strongly flattened; there is a pronounced
elongation in the east-west direction.

The X-ray emission from AWM~7 is similarly elongated;
Mohr et al. (1995) compute an axial ratio of $0.665\pm0.089$ at a position
angle of $-83^\circ \pm 7^\circ$ from the X-ray
flux within $7\farcm5$ of the cluster center, in perfect agreement with the
value of 0.67 computed by Dell'Antonio et al. (1995) for an isothermal
ellipsoid model density distribution.  Mohr et al. (1995) report X-ray 
ellipticities for 58 clusters; AWM~7 is more flattened than both the 
mean and median clusters, but there are 10 clusters that are more
flattened still. 
In our case it is difficult to measure the flattening
quantitatively from our optical sample for comparison with the X-ray
results because the completeness limit varies across our survey.

\begin{figure}[p]
\centerline{\epsfxsize=6.0in%
\epsffile{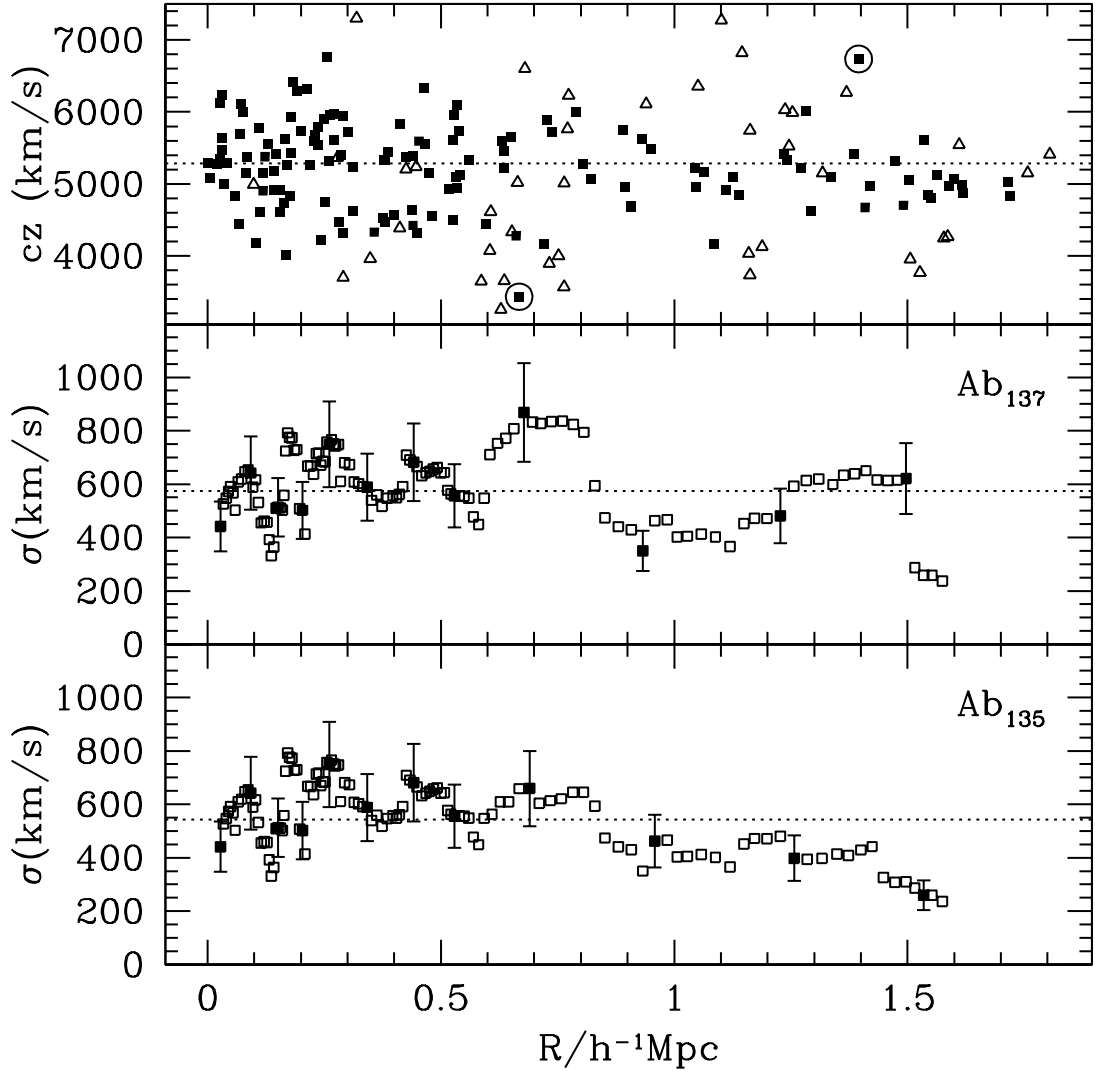}}
\caption{
 Top: Velocity as a function of projected distance from the cluster center.
 Ab galaxies are filled squares, Em galaxies are open triangles.  The two
 outliers are circled.
 Middle: Velocity dispersion profile for the 137 Ab galaxies in the cluster
 sample.  Every eleventh point is independent and is shown with 1$\sigma$ error
 bars.  The horizontal rule indicates these galaxies' overall dispersion
 of 574 \kms.  Bottom: Same as middle panel, but with two outliers removed.
 Horizontal rule is now at 543 \kms.
 }
\label{fig:posvel}
\end{figure}

Redshifts of cluster galaxies plotted against distance from the cluster
center define a trumpet-shaped locus delimited by caustics (Kaiser 1987).
Galaxies inside the caustics are bound to the cluster; those outside are
unbound.  We construct such a plot in Figure~\ref{fig:posvel} (top panel);
again, squares are Ab galaxies and triangles are Em galaxies.
Although one can discern by eye where the caustics probably lie,
they are in practice poorly defined.  Their accurate
determination requires many more galaxies than our sample contains (Diaferio
1999, Diaferio \& Geller 1997, Geller, Diaferio \& Kurtz 1999).

From the velocity histograms and the position-velocity diagrams it is clear
that the Em and Ab galaxies are not equivalent tracers of the cluster mass.
A friends-of-friends algorithm (Huchra \& Geller 1982) for finding 
substructure suggests that many of the Em galaxies are contained in
smaller subgroups distinct from the main cluster (not shown); these may
be accreting onto the cluster now, and, not yet being virialized, do not
follow the dark matter distribution as closely as the Ab galaxies.
There is considerable radial segregation of Ab and Em galaxies; the radial
distribution of Em galaxies is more uniform and has a greater mean and
median distance from the cluster center
than that of the Ab galaxies.  This behavior
is observed in other clusters (Mohr et al. 1996, Adami et al. 1998), in groups
(Mahdavi et al. 1999), and in the field as the
morphology-density relation (Dressler 1980).
Moreover, the velocity dispersion of the Em galaxies remains
large with increasing radius while that of the Ab galaxies declines
(see \S 5).  These considerations suggest that the Em galaxies are not 
virialized tracers of the underlying dark matter distribution.
Because our goal is 
an accurate determination of the mass profile, we exclude the Em galaxies
from further analysis, following Mohr et al. (1996).
We will revisit the issue of substructure in and around AWM~7
and other poor clusters in a subsequent paper.

Figure~\ref{fig:posvel} (middle panel) shows the velocity dispersion profile
constructed only from the Ab galaxies.  We sort the
galaxies by distance from the cluster center, then evaluate the velocity
dispersion in a sliding bin containing 11 galaxies; we plot
the dispersion as a function of the mean distance to the galaxies in the bin.
Neighboring points are strongly correlated, as they share 10 of their 11
galaxies; every 11th point (distinguished by error bars) is independent.  

Two large excursions from the overall trend dominate the profile.
Two outlier galaxies, one at projected distance
$R\sim0.7$~\hmpc\ and low velocity, the other
at $R\sim 1.4$~\hmpc\ and high velocity (circled in the top panel of the figure)
produce these excursions.  It is clear that these two galaxies
are responsible because the number of data points in each excursion equals
the width of the sliding bin used to compute the profile.
We therefore define a subset of the Ab sample, Ab$_{135}$, excluding these two
galaxies.  We shall henceforth refer to the
entire sample of 137 Ab galaxies as Ab$_{137}$.
The velocity dispersion profile for Ab$_{135}$ appears in the lower panel; the
overall dispersion of these 135 galaxies is $543^{+37}_{-31}$~\kms within
$R = 1.7$~\hmpc.
After some fluctuations near the core, the profile becomes smooth and 
shows the characteristic declining profile seen in poor groups by
Mahdavi et al. (1999) and in some $N$-body simulations
(e.g. Crone, Evrard \& Richstone 1994).
The decline in dispersion at small radii matches that of the ``type~C''
clusters of Girardi et al. (1998).  We revisit the issue of the cold core
in \S 6.

\section{Mass Profile}

We compute the mass profile of AWM~7 in two largely independent ways.
First, we assume that the mass profile follows the functional form
predicted by the hierarchical clustering simulations of Navarro, Frenk
\& White (1995,1996,1997).  This model has two parameters which we
derive separately from the observations under the assumption that the
Ab galaxies are virialized and that they trace the dark matter
distribution.  While this is a restrictive assumption, there is growing
evidence that the NFW profile is applicable beyond the hierarchical
clustering regime in which it was proposed.  Huss, Jain \& Steinmetz (1999)
find from high-resolution $N$-body simulations that virialized
dark matter halos with different formation histories nevertheless
evince similar density and velocity dispersion profiles, and that this
``universal'' density profile is well described by the functional form
of the NFW model.  The collapse history of the cluster seems immaterial
in determining the ultimate cluster profile.  More concretely, Geller,
Diaferio \& Kurtz (1999) show that this profile describes the Coma
cluster well for $R \lesssim 10$ \hmpc.  For comparison, we also fit
a non-singular isothermal sphere to AWM~7.

We also compute the mass profile from the virial theorem alone.  At small
radii, this method gives incorrect results if not corrected for the exclusion
of the outer cluster regions from the computation.
The mass profile can be corrected with a surface term, which depends on
both the mass and velocity dispersion profiles.  These last
two profiles are linked to the velocity anisotropy profile through the Jeans
equation (e.g. Binney \& Tremaine 1987), but the data are insufficient
to constrain the problem completely.  Usually one makes some assumptions
about one or more of the profiles and solves for the others (e.g. Binney \&
Mamon 1982, The \& White 1986, Binney \& Tremaine 1987, Girardi et al. 1998).
Here, we make these assumptions only for the computation of the surface term
correction to the virial mass estimator.

\subsection{The NFW model}

\subsubsection{Description}

The spherically symmetric NFW density distribution is
\begin{equation}
 \rho(r) = { {\delta_c \rho_c} \over
               {(r/r_c) (1+r/r_c)^2} }
\end{equation}
resulting in the enclosed mass profile
\begin{equation}
  M(<r) = 4 \pi \delta_c \rho_c r_c^3
           \left[ \log(1+r/r_c) - {{r/r_c}\over{1+r/r_c}} \right]
  \label{eq:nfwm}
\end{equation}
where $\rho_c$ is the critical density of the universe, and $\delta_c$ and
$r_c$ are the two parameters of the model, corresponding to the overall
normalization
and the core radius, respectively.  The radial scale in the context of
cluster simulations is often quoted in units of $r_{200}$, the radius where 
the mean cluster density has dropped to 200$\rho_c$.  NFW claim, on
the basis of numerical experiments, that this radius approximately separates
the virialized and infall regions (see also Cole \& Lacey 1996).
NFW define the ``concentration
parameter'' $c = r_{200}/r_c$.  A low value of $c$ arises if the core
radius is comparable with the overall extent of the cluster, that is, the
mass profile is not very concentrated toward the center.  NFW demonstrate 
that more massive clusters are less concentrated, with the concentration
parameter related to the overall normalization by
\begin{equation}
  \delta_c = {200 \over 3} {{c^3} \over {\log(1+c) - c/(1+c)}} .
\end{equation}
Because the enclosed mass of the NFW profile diverges logarithmically, it
is customary to quote $M_{200} \equiv 200 \rho_c (4\pi /3) r_{200}^3 $
as the cluster mass.

\subsubsection{Fitting the NFW Model}
\label{sec:maxlike}

We can derive $r_c$ observationally from the distribution of projected radii
of cluster galaxies.  The projected
surface number density of galaxies in the NFW model 
is given by
\begin{equation}
 \Sigma(\tilde R) \propto {{1-X(\tilde R)} \over {r_c^2 (\tilde R^2-1)}} 
\end{equation}
where $\tilde R = R/r_c$ is the projected radius in units of the core
radius, and
\begin{equation}
 X(\rt) = \cases{ (1-\rt^2)^{-1/2}{\rm sech}^{-1} \rt & if $\rt <    1 $\cr
                    (\rt^2-1)^{-1/2}     \sec ^{-1} \rt & if $\rt \geq 1.$\cr }
\end{equation}
Given the distribution of projected radii in our sample, we fit $r_c$
with a maximum-likelihood technique by finding the value of
$r_c$ that maximizes the probability
\begin{equation}
 {\cal L} = \prod_i \left [
            {{\rt _i\,\Sigma(\rt _i)} \over
                 {\int_0^{\rt _{\rm max}(r_c)}\rt\,\Sigma(\rt)\,d\rt}} 
            \right] ^{w_i}
 \label{eq:maxlike}
\end{equation}
of observing that particular distribution of projected radii, where
$w_i$ is a weight assigned to each galaxy to account for the non-uniform
magnitude limit of our survey.  We assign each galaxy one of two possible
weights as follows: galaxies within the central 0.51 \hmpc\ of the survey,
where it is 97\% complete to $m_R=17.2$, have $w=1.00$; galaxies with
$R_i > 0.51$~\hmpc, where the survey is complete to $m_R=15.5$, have $w=1.95$,
where 1.95 is the ratio of the number of galaxies with $m_R<17.2$ to the 
number with $m_R<15.5$ within 0.51~\hmpc\ of the cluster center.  This
correction assumes that the luminosity function is the same in the core
as in the outskirts of the cluster.  We exclude
galaxies fainter than $m_R=17.2$ within 0.51~\hmpc\ and fainter than
$m_R=15.5$ otherwise from the fit; there remain 53 galaxies with $R<0.51$~\hmpc\
and 70 with $R\ge0.51$~\hmpc.  Both outliers remain in the sample.
The denominator in Eq.~\ref{eq:maxlike}
serves to normalize the intrinsic probability of observing a
particular galaxy $i$ at projected radius $\rt _i$ (expressed in the numerator)
to the possible range of \mrt where it could be included in the sample.
The upper limit of the integral is a function
of $r_c$; $\rt _{\rm max}(r_c) = R_{\rm lim} / r_c$, where $R_{\rm lim}$ is
the limiting projected radius of the survey.  In practice, we use the largest
projected radius in the sample for $R_{\rm lim}$.  Table~\ref{tab:projrc}
lists the best-fit values of $r_c$ for the Ab$_{135}$ and Ab$_{137}$ subsamples,
computed from 121 and 123 galaxies, respectively, along with 1-, 2-,
and 3-$\sigma$ uncertainties, as determined by $\Delta(\log{\cal L}) =
0.5, 1.0, 1.5$.  There is little difference in the best-fit $r_c$ for the 
two samples; if the Em galaxies are included, the best-fit $r_c$ increases
by $\sim$50\%, because the Em galaxies are less centrally concentrated
(Fig.~\ref{fig:posvel}).  We compare the NFW model to the non-singular
isothermal sphere in Section~\ref{sec:nsis}.

\begin{figure}[p]
\centerline{\epsfxsize=6.0in%
\epsffile{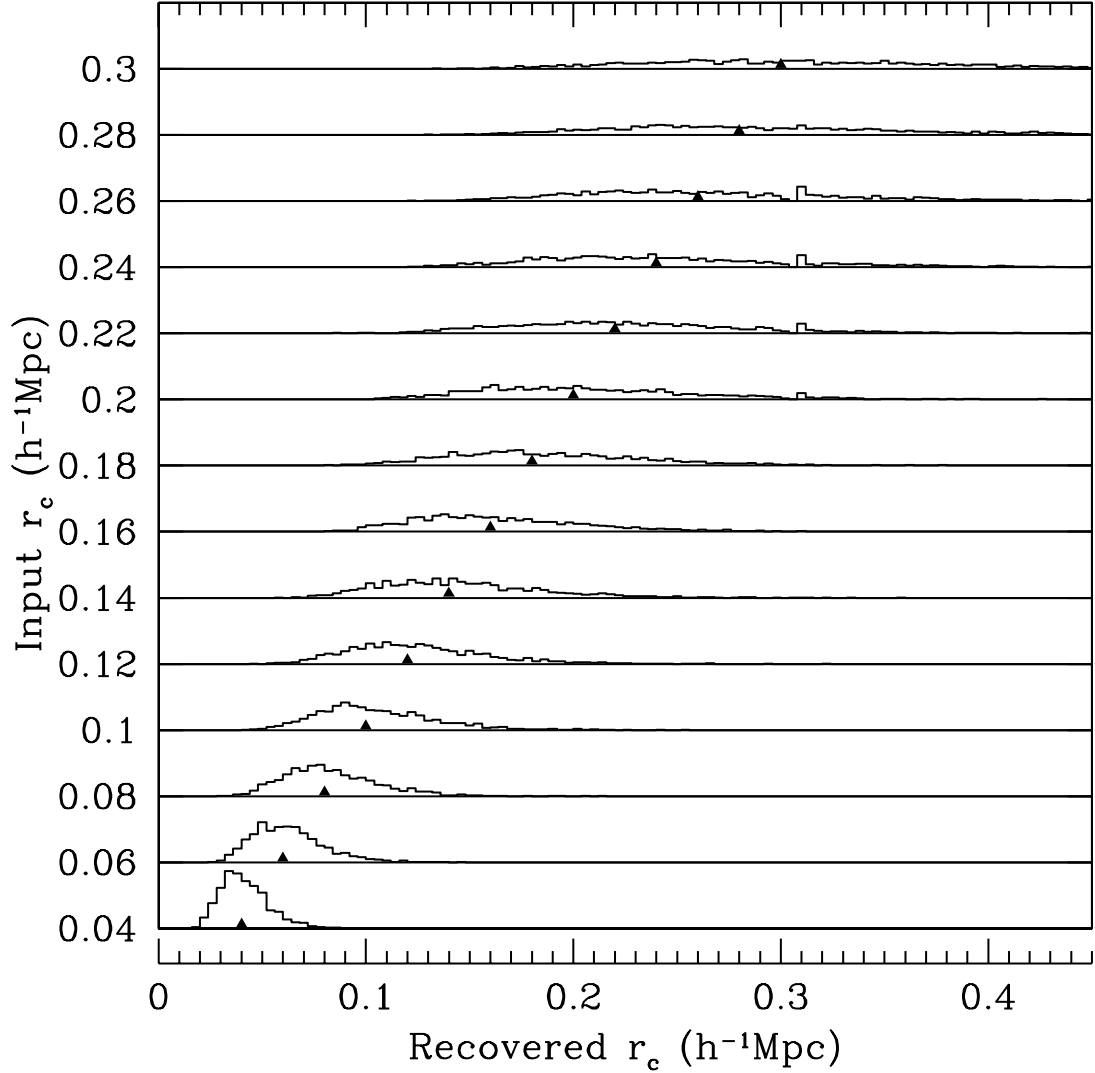}}
\caption{
 Histograms of recovered values of the NFW parameter $r_c$ from 2000 
 Monte Carlo realizations of sets of 123 projected radii drawn from NFW
 models with varying $r_c$.
 The histograms are all to the same scale, but for the highest input $r_c$
 may extend off the right edge of the plot.  The triangles in the histograms
 and the numbers along the vertical axis both indicate the input value of $r_c$.
 }
\label{fig:monterc}
\end{figure}

We use Monte Carlo simulations to assess the robustness of this method:
for selected values of $r_c$, we construct 2000 sets of 123 projected radii
drawn from the distribution of projected radii for an NFW
profile with that $r_c$; we reproduce the non-uniform magnitude limit of
our survey by only accepting galaxies with $R\ge0.51$~\hmpc\ with probability
1/1.95.  We compute the maximum-likelihood recovered $r_c$
using Eq.~\ref{eq:maxlike}.  Figure~\ref{fig:monterc} shows the results
for $r_c$ in the range 0.04 -- 0.30~\hmpc.
The median recovered $r_c$ is within 0.002 of the input $r_c$ in all but
one case; however, the distributions of the recovered $r_c$ are broad.
The distribution of recovered $r_c$ is narrowest for small $r_c$, and 
broadens as $r_c$ increases, because (1) the peak in the probability
distribution of projected radii becomes less pronounced as $r_c$ increases,
and because (2) we sample a smaller dynamic range of \mrt as $r_c$ increases
(because the limiting projected radius $R_{\rm lim}$ is constant).
The distribution of the ratio of recovered to input $r_c$ is largely
independent of input $r_c$, however.  The lower and upper $1\sigma$
bounds (defined as the values of $r_c$ enclosing 34.2\% of the distribution
on either side of the median) are typically displaced $-$25\% and $+$30\% from
the true $r_c$, respectively.  Increasing the sample size
in the simulations to 500 radii (not shown) reduces
the spread in the distribution of recovered $r_c$ by a factor of $\sim$2.

Knowing $r_c$, the NFW model would be completely specified if we also knew
$r_{200}$, but this quantity is difficult to derive observationally.
We can arrive at an estimate if we assume that the virial theorem holds,
so that $M_v \equiv 3 \sigma_p^2 r_v G^{-1}$, where $G$ is the
gravitational constant, $r_v$ is the virial radius, and $\sigma_p$ is the
global line-of-sight cluster velocity dispersion.
Then
\begin{equation}
 {{M_{200}r_v}\over{M_v r_{200}^3}} = 100 H_0^2 {{(1+z)^3}\over{3 \sigma_p^2}}
 \label {eq:r200}
\end{equation}
where we have used $\rho_c(z) = 3 H_0^2 (1+z)^3  / ( 8\pi G )$ for
$\Omega = 1$, and the definition of $M_{200}$.
The right-hand side of this equation is observationally determined, and
substituting for $M_v$ and $M_{200}$ using Eq.~\ref{eq:nfwm},
the left-hand side becomes a function of $r_c$, $r_v$, and $r_{200}$ which we
solve for $r_{200}$.  We fit $r_c$ from the distribution of projected
radii, and approximate $r_v$ by
\begin{equation}
 r_v \sim {\pi \over 2} {{N(N-1)}\over{\Sigma_{i}\Sigma_{i<j} R_{ij}^{-1}}} 
 \label {eq:rv}
\end{equation}
where $N$ is the number of galaxies in the system, and $R_{ij}$ is the
projected separation between galaxies $i$ and $j$ (Binney \& Tremaine 1987).
This estimator yields $r_v \sim 1.30$ \hmpc\ for the Ab$_{137}$ sample, and 
$r_v \sim 1.28$ \hmpc\ for the Ab$_{135}$ sample.

The value of $r_{200}$ that solves Eq.~\ref{eq:r200} is fairly insensitive
to the value of
$r_c$; in the regime of interest, where $ 500 < \sigma_p < 600$ \kms, 
$r_{200}$ varies $\sim$3\% as $r_c$ ranges from 0.12 to 0.30~\hmpc\ for
any constant $\sigma_p$.  The derived $r_{200}$ is more sensitive to
$\sigma_p$, varying $\sim$15\% over the range $ 500 < \sigma_p < 600$ \kms
with constant $r_c$.  For the $r_v$ obtained from Eq.~\ref{eq:rv}, the
values of $r_{200}$ are 1.02 and 0.97 \hmpc\ for the Ab$_{137}$ and
Ab$_{135}$ samples, respectively (Table~\ref{tab:nfwpar}).
The corresponding value of the concentration
parameter $c$, 4.6 in both cases, falls in the middle of the range 2.3--7.7
seen by Carlberg et al. (1997) for rich systems, although the errors are
large in both cases.  Comfortingly, the values are also smaller
than the $c \sim 7.5$ that Mahdavi et al. (1999) find for their
lower-mass poor groups.  For AWM~7, $M_{200}$ is
2.47 and $2.13 \times 10^{14}$ $h^{-1} M_\odot$ for the Ab$_{137}$ and 
Ab$_{135}$ samples, respectively.  Here the larger mass is associated
with a formally slightly smaller value of $c$ (4.56 vs. 4.61), in keeping with
the general trend seen by NFW, but this consistency may be accidental,
considering the uncertainties in the 
computed values of $c$. 
It is interesting to note that two outliers (out of a sample of 137) can
alter the recovered mass by $\sim$15\%.

Because the cluster extends beyond $r_{200}$, $M_{200}$ in fact underestimates
the total mass.  Extrapolating the NFW mass profile to 1.7 \hmpc\ (the maximum
radial extent of the Ab sample) with Eq.~\ref{eq:nfwm}
yields 3.49 and 3.01 $\times 10^{14}\, h^{-1} M_\odot$ for the Ab$_{137}$ and
Ab$_{135}$ samples, respectively.

\subsection{Comparison to the Non-Singular Isothermal Sphere}
\label{sec:nsis}

\begin{figure}[p]
\centerline{\epsfxsize=6.0in%
\epsffile{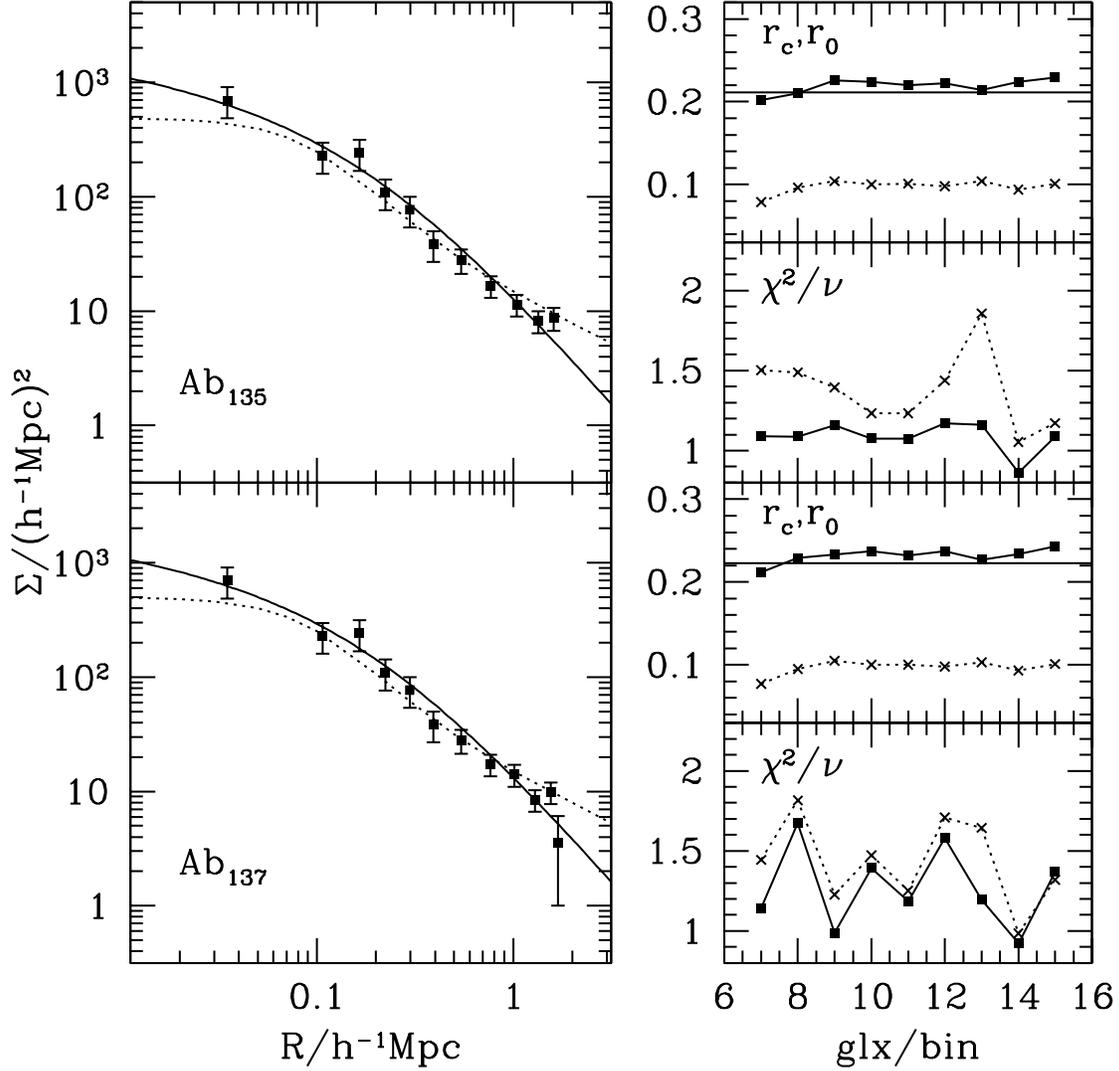}}
\caption{
 Comparison of the actual binned projected surface number density of galaxies
 to the NFW (solid) and non-singular isothermal sphere models (NSIS, dotted).
 The top half of
 the figure refers to the Ab$_{135}$ sample; the bottom half to the Ab$_{137}$
 sample.  Left: the actual surface density profile is shown as the dots with
 error bars; there are 11 galaxies per bin.  The curves show the best-fit
 NFW and NSIS models.  Right: dependence
 of the fits on the binning.  The behavior of the best-fit value of $r_c$ 
 (for the NFW model) or $r_0$ (for NSIS) and of $\chi^2/\nu$ is shown as 
 a function of the binning.  The horizontal rules indicate the 
 maximum-likelihood values of $r_c$ for the two samples.
 }
\label{fig:projcomp}
\end{figure}

The non-singular isothermal sphere (NSIS) arises from setting finite boundary
conditions at zero radius to the same differential equation that gives
rise to the singular isothermal sphere (e.g. Binney \& Tremaine 1987,
\S 4.4).  It is characterized by a density normalization $\rho_0$ and a
core radius $r_0$ at which the the projected density falls to 0.5013 of 
its central value.  In Fig.~\ref{fig:projcomp} we compare the fit of
the projected surface number density from the NSIS and NFW models to the 
data.  Although the maximum-likelihood technique of \S\ref{sec:maxlike}
makes optimal use of the data, it provides no independent assessment of
goodness-of-fit.  Here, although with some loss of information,
we bin the observed surface number density for comparison with the models;
we retain the same weighting scheme as in the maximum-likelihood 
calculation.

The left side of Fig.~\ref{fig:projcomp} shows the observed profile
(points with Poisson error bars), the NFW fit (solid line),
and the NSIS fit (dotted line) for both the Ab$_{135}$ (top panel)
and Ab$_{135}$ (bottom) samples.  The right side of the figure illustrates
how the best-fit values of $r_c$, $r_0$, and $\chi^2/\nu$ vary with the
choice of binning; the profiles and fits are shown for 11 galaxy
equivalents per bin (galaxies beyond 0.51~\hmpc\ count as 1.95 equivalents).  

For different binnings 
the quality of the fits of both the NFW and NSIS models, as measured by
$\chi^2/\nu$, varies much less for the \abf\ sample
than for the \abs\ sample.  The best-fit values of $r_c$ and $r_0$
are quite stable for both samples.  The decreased sensitivity to the binning
provides further justification for excluding the two outlier galaxies.  For
both samples, the NFW model fits with a lower $\chi^2/\nu$ than the NSIS
model, independent of the binning (except in one case).  The superiority
of the NFW model is more consistent for the \abf\ sample,
although this superiority is not overwhelming; for \abf, NFW fits with
$\chi^2/\nu = 1.075$, whereas the NSIS fits with $\chi^2/\nu = 1.233$.  For
\abs, the respective $\chi^2/\nu$ for NFW and NSIS are 1.184 and 1.252
(all for 11 galaxies per bin).  There is thus some formal statistical
justification for using the NFW model over the NSIS, but 
an unambiguous determination would require a survey out to
2 or 3~\hmpc\ (cf. Geller, Diaferio \& Kurtz 1999).

With 11 galaxies per bin, the best-fit values of $r_c$ are within 5\% of
the maximum-likelihood values for both samples, and well within the 
1-$\sigma$ range for all binnings, indicating that the loss of information
inherent in the binning is not severe.

\subsection{The Virial Mass Profile}

We now compute the mass profile of AWM~7 from the virial estimator appropriate
for the case of galaxies embedded in a diffuse distribution of dark matter,
under the added assumption that light traces mass.  The appropriate estimator
(Binney \& Tremaine 1987) is 
\begin{equation}
 M_{\rm est} = {{3\pi N} \over {2G}}
                 {{\Sigma_i v_i^2 } \over
                  {\Sigma_i \Sigma_{j<i} |{\bf R}_i-{\bf R}_j|^{-1} } } 
             = {{3 r_v}\over{G}} {{\Sigma_i v_i^2}\over{N-1}}
\label{eq:vtmass}
\end{equation}
where $v_i$ is the radial velocity relative to the cluster mean, and ${\bf R}_i$
is the projected radius vector relative to the center of the cluster.
This estimator assumes that the galaxies are in dynamical equilibrium within
the cluster potential, that galaxies trace the dark matter distribution,
and that the entire cluster has been observed.  These
assumptions do not apply to the Em galaxy population; indeed, when we apply
this prescription to the Em galaxies alone, the resultant mass estimate is a 
factor of $\sim$4 greater than for the Ab galaxies.  To correct for
the failure to include the entire angular extent of the cluster
in our sample, we correct the mass estimate with a surface term $C(b)$
(The \& White 1986) that depends on the limiting radius $b$ of the 
observations.  The corrected mass is
\begin{equation}
 M_{cv}(<b) = M_{\rm est}[1-C(b)]
            = M_{\rm est} \left[ 1 -
      4\pi b^3 { {\rho(b)} \over {\int_0^b 4\pi r^2 \rho(r)\, dr} } \,
               { {\sigma^2_r(b)} \over {\vphantom{\int_0^b}\sigma^2(<b)} }
    \right] 
 \label{eq:surfterm}
\end{equation}
where $\sigma(<b)$ denotes the integrated velocity dispersion within 
the limiting radius $b$ (e.g. Girardi et al. 1998). 

The surface term incorporates information about the velocity dispersion
profile and the density profile; the former is 
measurable from the data, but some assumption must be made
about the form of the density profile --- in order to correct that very profile.
The \& White (1986) and Girardi et al. (1998) posit a functional form
for the mass profile, which they then fit to the data on the basis of
the projected number density of galaxies.  Here we assume an underlying
NFW mass profile.

\subsubsection{Behavior of The Surface Term}

The radial dependence in the NFW model of the surface term in
Eq.~\ref{eq:surfterm} is
\begin{equation}
   C_{\rm NFW}(r) = 
   {{u^2}\over{(1+u)^2}} \left[ \log(1+u) - {{u}\over{1+u}} \right] ^{-1}
    \left[ {{\sigma_r(r)} \over {\sigma(<r)}} \right] ^2
\end{equation}
where $u = r/r_c$.  The last term involves both the overall velocity dispersion
$\sigma$ and the radial velocity dispersion $\sigma_r$, and is thus a function
of the anisotropy parameter 
$\beta = 1 - \sigma_\theta^2 / \sigma_r^2$, which is itself a function of $r$.
Now $\sigma^2 = \sigma_r^2 + \sigma_\theta^2 + \sigma_\phi^2$; under the
assumption of spherical symmetry the last two terms are equal,
so $\sigma^2 = (3-2\beta)\sigma_r^2$, which we substitute into the denominator
of the last term for computation.  The anisotropy and velocity dispersion
profiles are not independent; they are coupled via the Jeans equation.  
Girardi et al. (1998) find that for their mass model
a constant anisotropy profile $\beta(r) = 0$
produces a velocity dispersion profile that peaks at $\sim$0.1~$r_v$, dips
slightly in the core, and decreases gradually at large radii; the simple
functional form $\beta(r) = -k/r$, where $k$ is a constant, produces a 
velocity dispersion profile with a more severe dip in the core.  In this case,
the orbits in the core become purely tangential as $\beta \rightarrow -\infty$.

\begin{figure}[p]
\centerline{\epsfxsize=6.0in%
\epsffile{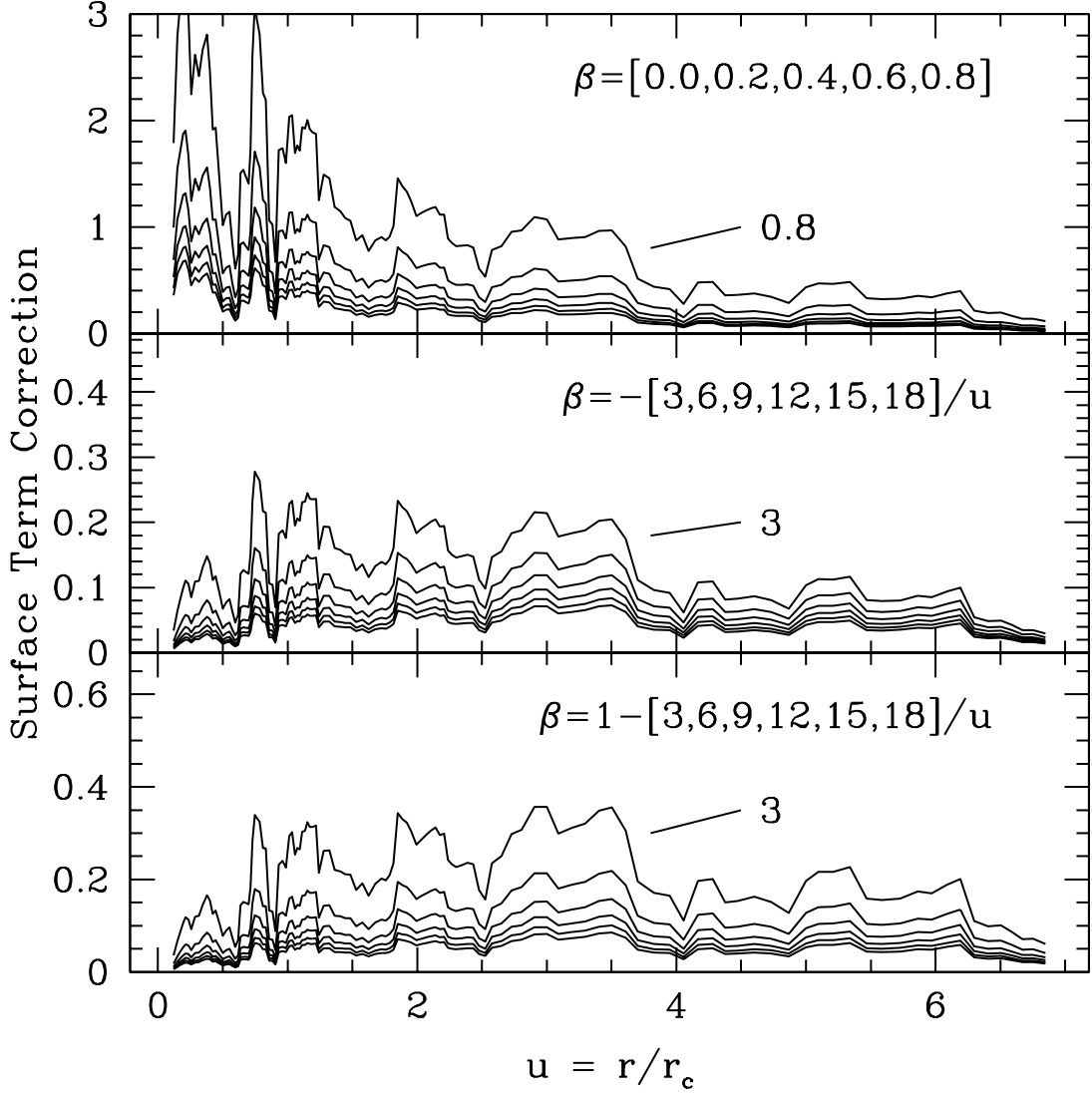}}
\caption{
 Behavior of the surface term correction to the virial mass estimator
 as given by Eq.\ref{eq:surfterm} assuming an underlying NFW mass profile
 for three models of the anisotropy profile $\beta(r)$.  The assumed NFW
 profile uses $r_c = 0.211$ (the best-fit value for the Ab$_{135}$ sample),
 and we use the observed velocity dispersion profile.  Top panel: 
 $\beta(r) = k$; middle: $\beta(r) = -k/r$; bottom: $\beta(r) =
 1 - k/r$, where $k$ is a constant in each case.  There is one curve for
 each value of $k$ listed in the label, with one of the extreme curves
 labeled for reference.
}
\label{fig:surfterm}
\end{figure}

Figure~\ref{fig:surfterm} shows the behavior of the surface term for an
NFW mass distribution and the observed velocity dispersion profile
(Fig.~\ref{fig:posvel}) under
the assumption of three different anisotropy profiles:
constant (top panel), $\beta(r) = -k/r$ (middle), and $\beta(r) = 1 -k/r$
(bottom).  The multiple curves in each panel correspond to the values of 
$\beta$ (top) or $k$ (middle and bottom) enclosed in brackets.  One of the
extreme curves is labeled in each case for reference.  The values of $k$
are chosen to bracket the value inferred from the profiles in Fig.~3 of
Girardi et al. (1998).  Note that the three panels are not drawn to the
same scale.  Because these classes of $\beta$ profile are derived from
fits to the mass model of Girardi et al. (1998), they are not a priori
strictly valid for the NFW profile; nevertheless, it is reasonable to
adopt them here as representative of the general forms that the profiles
may have.

By definition, the surface term must be everywhere less than unity (to
ensure a positive mass).  The results shown for constant large values
of $\beta$ are thus clearly unphysical.  At small radii, the data
are too noisy for all our assumptions to hold; a proper treatment would 
require a full solution of the Jeans equation.  It is unlikely that this 
treatment would improve the results very much, however, because of the 
intrinsic noisiness and discreteness of the data.  At larger radii, the 
variations in the surface term become less pronounced, and the surface
term in general becomes less important, decreasing to $\lesssim 5\%$ for
each profile at the outer limit of our sample.  Sufficiently far from the
core, then, the virial mass estimator provides a robust estimation of the 
enclosed cluster mass, but at small radii the surface correction is too
variable to result in a reliable mass profile.  The virial mass estimator
thus seems to be a good tool for estimating total cluster masses, but it is 
not well suited to the determination of mass profiles in cluster cores,
where the surface term correction is most important but least well determined.

A reliable description
of the surface term at small radii may only be possible from cluster
simulations in which the dark matter density and velocity distributions
are fully known to the experimenter.  In that case, empirical surface 
term profiles can be built from the data and compared to reconstructions
that an observer would generate under various assumptions of
the anisotropy profile.

\subsubsection{The Virial Mass Profile}

\begin{figure}[p]
\centerline{\epsfxsize=6.0in%
\epsffile{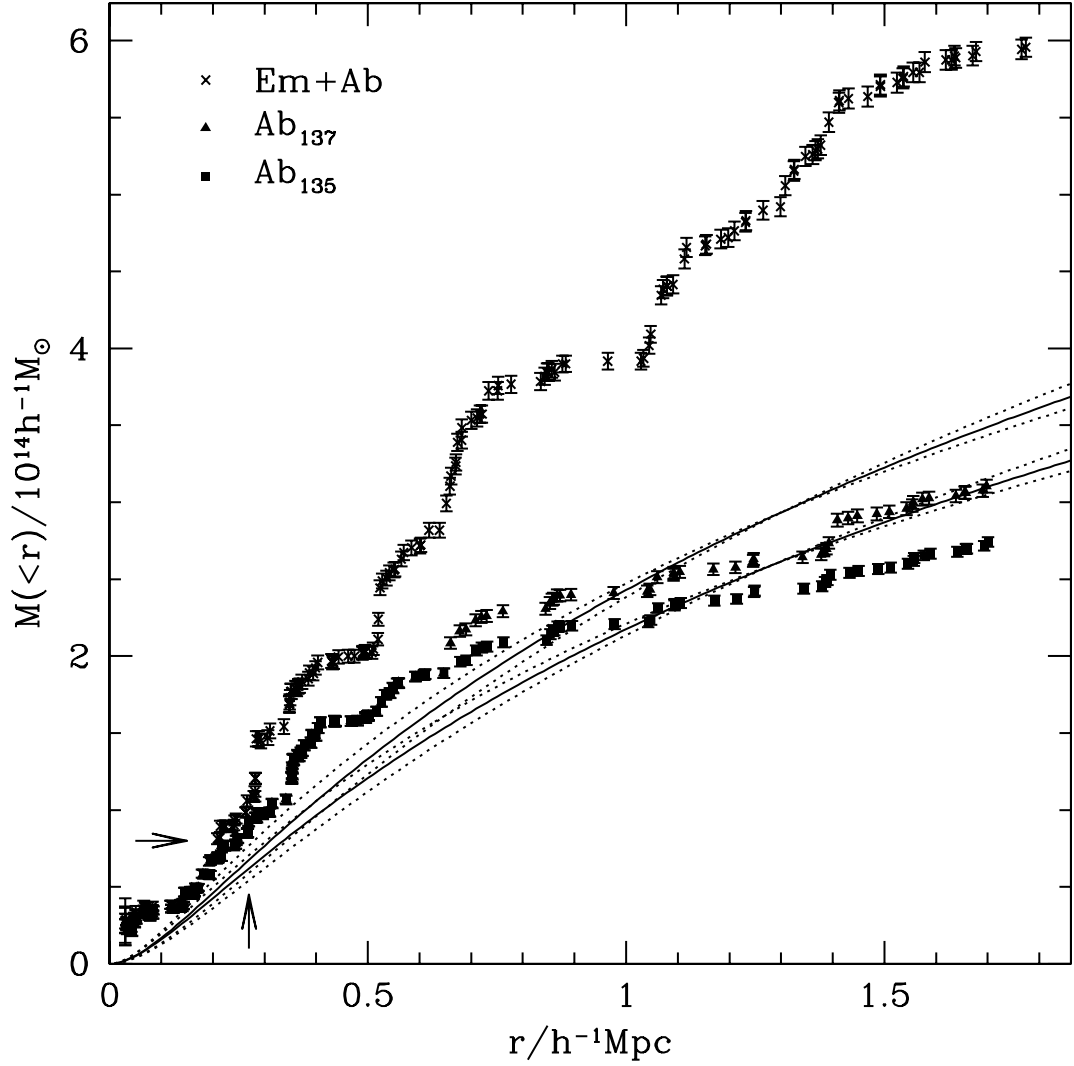}}
\caption{
 Enclosed mass profile as determined from the virial theorem, for 
 Em, Ab$_{137}$, and Ab$_{135}$ samples.  Error bars show the formal
 bootstrap errors (see text). The solid curves are the independently
 derived NFW profiles tabulated in Table~\ref{tab:nfwpar}; the dotted
 curves are the profiles corresponding to the 1$\sigma$ ranges of $r_c$
 from Table~\ref{tab:projrc} with $r_{200}$ recomputed from
 Eq.~\ref{eq:r200}.  The arrows point to the X-ray mass of Dell'Antonio
 et al. (1995).
}
\label{fig:vtmass}
\end{figure}

Figure~\ref{fig:vtmass} shows the virial mass profile (Eq.~\ref{eq:vtmass}),
with no surface term correction, as derived from three
subsets of our galaxy sample: the entire sample (crosses), the Ab$_{137}$
sample (triangles), and the Ab$_{135}$ sample (squares).  The entire (Em+Ab)
sample is included only to illustrate that the inclusion of the emission-line
galaxies with their larger velocity dispersion skews the results towards
higher mass.  Indeed, this difference is 
evident in the two Ab profiles as well: they agree up to the first outlier,
at which point the Ab$_{137}$ mass profile jumps; the two Ab profiles then
grow more or less in parallel until the next outlier.
We compute uncorrected enclosed masses within 1.7 \hmpc\ of
3.10 and $2.74 \times 10^{14}$ $h^{-1} M_\odot$
from the Ab$_{137}$ and Ab$_{135}$ samples, respectively.  The surface
term correction at this radius is on the order of 2--5\%, depending on
the anisotropy.

We estimate the error in the enclosed mass profile
by the statistical bootstrap method (Diaconis \& Efron 1983):
for the $N$ galaxies enclosed within a given
projected radius, we compute the $N$ possible enclosed masses from
all subsets of $N-1$ galaxies.  We take the standard deviation of the resultant
set of masses as the formal error in the estimated enclosed mass.  The formal
errors become quite small, on the order of 3\%, at large radii.  When 
the number of galaxies is large and the distribution of
velocities is well-behaved, the random rearrangement of velocities
has little effect on the mass estimate; thus, the distribution
of the bootstrapped masses is narrow.  The actual error is in fact larger,
as demonstrated by the sensitivity to the outliers.
The error in the enclosed virial mass is $\sim$15\%.

The solid smooth
curves indicate the NFW mass profiles we fit previously, and the dotted
lines indicate the profiles corresponding to the 1$\sigma$ ranges of 
$r_c$ from Table~\ref{tab:projrc} with $r_{200}$ recomputed from
Eq.~\ref{eq:r200}.  The agreement with the virial mass is 
quite good, to within $\sim$15\% at 1.7 \hmpc\ for both Ab samples, with
the virial mass the smaller of the two in each case.  The agreement along
the profile in general is reasonable given the uncertainty in the surface term.
The discrepancy is greatest near $R\sim 0.5$~\hmpc, corresponding to
$r/r_c \sim 2$, where the surface term correction can be on the order
of 20\% in the proper direction; at the virial radius the masses are
in perfect agreement, and at larger radii the NFW mass profiles overtake
the virial mass profiles.
Because $r_{200}$ is insensitive to $r_c$, the variation in $r_c$ results
in little difference in the mass profile at large radii, but of course
has a greater effect on scales of a few times $r_c$.  Table~\ref{tab:masscomp}
compares the NFW and virial masses.  Based on the sensitivity of the
virial mass to the outliers and on the effect of the surface term, we assign
an error of $0.5\times10^{14}\, h^{-1}\,$M$_\odot$ to our mass estimate.

Our mass profile is in agreement with the ROSAT X-ray mass of Dell'Antonio
et al. (1995), who derive $M(<0.27) = 0.8\times 10^{14}$~M$_\odot$
(indicated by the arrows in Fig.~\ref{fig:vtmass}), and is
within the allowable range of mass profiles determined from ROSAT data by 
Neumann \& B\"{o}hringer (1995) and from ASCA data by
Loewenstein \& Mushotzky (1996) and Markevitch \& Vikhlinin (1997).
The X-ray masses are all quite uncertain and extend only to
$R \sim 0.5$~\hmpc.  Koranyi et al. (1998) illustrate the range of mass
profiles from X-ray determinations; the range is so broad as to include
even the virial mass profile computed from emission-line galaxies; any
plausible kinematic mass profile is thus likely to be consistent with
the current X-ray limits on the mass.
The increased sensitivity of the {\em Chandra} Observatory will allow
X-ray mass profiles to be determined with more precision and to larger radii.

\section{Discussion}

Galaxy clusters display a range of velocity dispersion profiles ( e.g.
den Hartog \& Katgert 1996, Girardi et al. 1998).  Some are peaked at
the center, others are more or less flat, still others rise with
increasing radius near the center.  This variety suggests that
clusters today are
still in a variety of dynamical states, and that indiscriminately
combining a set of clusters to improve the statistics can easily
lead to incorrect results if clusters with different kinematics are
averaged together.  Even within a single cluster, it is important to
discriminate between the generally red, virialized population and the
newly infalling bluer galaxies; including the latter in steady-state
kinematical analyses will artificially inflate the mass estimate; we
suggest that this effect accounts for the mass of $5.77^{+1.80}_{-1.50}$
$h^{-1} M_\odot$ derived for AWM~7 by Girardi et al. (1998).

In principle, the non-virialized Em galaxies can be used as tracers of
the escape velocity to arrive at an independent mass measurement
(Diaferio \& Geller 1997, Diaferio 1999).  In practice, there are not
enough Em galaxies in our sample
for this method to yield a robust mass estimate; although the formal
mass returned by this method for AWM~7 (kindly computed for us by
A. Diaferio) is comparable to the ones we derive here, the fractional
error in that mass is $\sim$0.9.

Our enlarged sample confirms the presence of a cold core in the cluster
(Koranyi et al. 1998); although still noisy near the cluster center, the
velocity dispersion within a projected radius of $\sim$0.2~\hmpc\ is lower
than in the range 0.2--0.5~\hmpc.  Beyond $\sim$1~\hmpc\ the velocity
dispersion drops below the range in the core.  Abell~576 (Mohr et al.
1996), a similarly well-sampled cluster, has a more pronounced cold core
but a rising absorption-line velocity dispersion profile in the range
0.4--1.2~\hmpc. 
A simple dynamical argument suggests that cold cores arise naturally from
the NFW density profile.  Assuming virial equilibrium to hold in the core,
$\sigma^2 \propto GM(<r)/r$.  Modeling the density profile at small radii
as a power law $\rho(r) \propto r^{-\alpha}$ yields $M(<r) \propto 
r^{3-\alpha}$, so $\sigma \propto r^{1-\alpha/2}$.  For the NFW profile
$\alpha = 1$ in the core, resulting in a rising velocity dispersion profile.

Because the velocity dispersion profile, mass profile, and orbital
anisotropy profile are coupled through the Jeans equation, it is 
difficult to derive a mass profile without making some assumptions
about the form of one or more of these functions along the way.  The
projected velocity dispersion profile is directly observable, but it is
sensitive to outliers and substructure, and requires either extensive
smoothing or a large number of galaxies for a robust determination.
The anisotropy profile is also difficult to constrain,
with some evidence for a variety of profiles possible in clusters
(Girardi et al. 1998).  
The mass profile can be estimated from X-ray observations, but
these tend to have few data points, large errors, and cover only the 
inner regions of clusters; consequently, they can be fit by a wide
range of models, resulting in large uncertainties in mass when 
extrapolated to large cluster radii.

It seems difficult to avoid circular logic in computing the surface 
term, as it depends on both the anisotropy profile and the density distribution
that one is trying to determine.  One can make a self-consistency argument,
as we do here, to verify that the assumed density profile used to correct
the virial mass profile agrees with the virial profile so corrected; if the 
surface term were small, this circularity would be less troublesome, because
the assumed density profile would affect the virial mass profile only
through that small correction.  But as Fig.~\ref{fig:surfterm} illustrates,
the surface term correction is often significant, particularly at small
radii.  Not only is the correction large, but it is also uncertain, 
sensitive to uncertainty in the anisotropy profile alone.
The coupling between the mass profile and the surface term is thus
quite strong; even with the assumption of a density
profile (which in principle renders the whole exercise redundant because
the mass profile is then already known), the surface term correction
is too poorly constrained to produce a reliable mass profile.
Only at the edges of the cluster does the surface term correction become
reliable, but by then it is no longer very important.  Accurate 
determination of cluster masses from the virial estimator, then, requires
broad angular coverage; dense sampling of the core is not enough.

\section{Conclusion}

We derive the mass profile of AWM~7 in two independent ways from an
optical sample of 179 galaxies with redshifts within 1.7 \hmpc\ of the
cluster center; we use only the 137 
absorption-line galaxies (and a 135-element subset of them) in the 
analysis.  One method uses the surface number density and global
velocity dispersion; the other uses individual velocities and
pairwise distances.  Both assume and underlying NFW profile.
The two methods yield remarkably consistent
mass estimates, particularly at the outer limit of our sample;
the enclosed mass within a projected radius of 1.7 \hmpc\ agrees
within 15\% for both samples.  The mass estimates are sensitive
to the inclusion of two outliers in the sample; both the NFW and
virial masses increase by $\sim$15\% when they are included.  Including the 
Em galaxies, with their larger dispersion, in the calculation results
in a virial mass larger by a factor of two.  The cluster mass within
1.7~\hmpc\ is $\sim$$(3\pm0.5)\times 10^{14}$~$h^{-1}$M$_\odot$.

The surface term correction to the virial mass estimator
is sensitive to noise in the velocity dispersion profile and to the 
assumed velocity anisotropy profile.  The surface term at small radii
is too unreliable for the virial mass estimator to be used for measuring
mass profiles; at large radii, where the surface term is more certain,
it is no longer an important correction to the enclosed mass.  The 
virial mass estimator is thus reliable for measuring total cluster masses,
but not enclosed mass profiles; to overcome the limitations of the 
surface term at small radii, cluster redshift samples must have broad
angular coverage.  Dense sampling of cluster cores can reduce the scatter
in the surface term if there are enough galaxies in the core, but our
experience with AWM~7 (22 cluster members out of 154 Hydra redshifts)
suggests that telescope time is better spent probing the outskirts of
the cluster.

Untangling the relation between the anisotropy profile, velocity dispersion
profile, and density profile in cluster cores may ultimately require 
high-resolution simulations that track both dark matter particles and,
once they have formed, galaxies.


\acknowledgments

We thank Jim Peters, Perry Berlind, and Mike Calkins for obtaining many
of the FAST spectra reported herein, and Susan Tokarz for preliminary
reductions.  We thank  Mike Kurtz and Doug Mink for advice in reducing the
Hydra spectra, Norman Grogin and Andisheh Mahdavi for useful discussions,
and Antonaldo Diaferio for lucid discussions and for running our data
through his caustic-finding algorithm.

This work is supported by the Smithsonian Institution.


\clearpage

\begin{deluxetable}{rrrrrrrr}
\tablenum{1}
\tablecolumns{8}
\tablecaption{Redshifts of Target Galaxies}
\tablehead{
 \colhead{$\alpha_{2000}$} &
 \colhead{$\delta_{2000}$} &
 \colhead{$cz$} &
 \colhead{$\sigma_{cz}$}  &
 \colhead{$\alpha_{2000}$} &
 \colhead{$\delta_{2000}$} &
 \colhead{$cz$} &
 \colhead{$\sigma_{cz}$}             \\
 \colhead{} & \colhead{} & \colhead{[km s$^{-1}$]} & \colhead{[km s$^{-1}$]} &
 \colhead{} & \colhead{} & \colhead{[km s$^{-1}$]} & \colhead{[km s$^{-1}$]} 
}
\startdata
2 45 12.81 & 41 44 04.57 & 4970 & 49 & 2 54 44.00 & 41 39 18.00 & 5372 & 25 \nl
2 45 31.90 & 41 14 34.34 & 5127 & 51 & 2 54 44.10 & 41 52 08.80 & 5974 & 24 \nl
2 45 36.34 & 42 09 27.32 & 5063 & 22 & 2 54 44.60 & 41 31 41.40 & 4450 & 22 \nl
2 45 59.29 & 41 19 18.70 & 5316 & 19 & 2 54 47.60 & 41 18 50.00 & 5904 & 40 \nl
2 46 24.61 & 41 16 36.62 & 4672 & 34 & 2 54 48.10 & 41 24 34.00 & 5622 & 16 \nl
2 46 57.12 & 41 59 15.58 & 5098 & 38 & 2 54 49.80 & 41 37 25.39 & 5994 & 38 \nl
2 46 58.43 & 41 46 33.28 & 4628 & 23 & 2 54 55.30 & 41 24 15.41 & 5423 & 42 \nl
2 47 40.54 & 40 29 41.96 & 5607 & 32 & 2 54 56.30 & 41 29 46.79 & 4615 & 41 \nl
2 47 57.73 & 40 18 29.92 & 4985 & 22 & 2 55 00.60 & 42 08 32.00 & 5613 & 26 \nl
2 47 58.14 & 41 19 23.95 & 4855 & 47 & 2 55 01.80 & 41 26 51.30 & 4918 & 51 \nl
2 48 17.12 & 41 09 49.21 & 5090 & 30 & 2 55 02.00 & 41 31 30.40 & 5781 & 22 \nl
2 48 22.03 & 41 36 47.23 & 5217 & 23 & 2 55 02.13 & 41 03 49.45 & 4556 & 41 \nl
2 48 49.65 & 40 40 46.42 & 5219 & 19 & 2 55 02.40 & 41 36 24.10 & 4175 & 18 \nl
2 48 59.01 & 41 02 21.12 & 5169 & 43 & 2 55 04.80 & 42 15 45.80 & 5223 & 23 \nl
2 49 09.34 & 41 34 57.79 & 4686 & 24 & 2 55 05.53 & 41 38 28.90 & 5377 & 32 \nl
2 49 09.55 & 41 34 47.71 & 4698 & 22 & 2 55 06.00 & 41 44 00.38 & 5926 & 53 \nl
2 49 20.60 & 40 53 17.02 & 4160 & 31 & 2 55 08.20 & 41 29 17.09 & 4917 & 40 \nl
2 49 40.90 & 41 03 17.00 & 5488 & 31 & 2 55 14.10 & 41 47 29.80 & 5797 & 52 \nl
2 49 48.10 & 41 27 45.50 & 5279 & 17 & 2 55 16.60 & 41 20 20.51 & 5324 & 31 \nl
2 49 51.00 & 41 34 03.70 & 5999 & 29 & 2 55 17.33 & 41 34 49.01 & 5177 & 25 \nl
2 50 03.48 & 43 00 09.54 & 5052 & 45 & 2 55 19.01 & 41 34 28.02 & 5411 & 21 \nl
2 50 52.80 & 41 45 52.70 & 5449 & 62 & 2 55 20.97 & 41 38 33.90 & 4742 & 38 \nl
2 51 17.70 & 41 09 18.40 & 3433 & 21 & 2 55 28.58 & 41 44 40.95 & 5685 & 18 \nl
2 51 18.60 & 41 33 55.10 & 5733 & 34 & 2 55 32.50 & 41 29 53.30 & 5735 & 49 \nl
2 51 43.22 & 42 49 43.39 & 5420 & 33 & 2 55 38.48 & 41 27 58.40 & 5600 & 63 \nl
2 51 52.26 & 41 30 38.88 & 4324 & 68 & 2 55 43.55 & 41 28 37.70 & 5538 & 66 \nl
2 52 07.90 & 41 34 45.90 & 4572 & 17 & 2 55 44.03 & 41 07 48.05 & 5559 & 17 \nl
2 52 15.50 & 41 23 55.10 & 5835 & 24 & 2 55 48.00 & 41 59 20.11 & 4423 & 23 \nl
2 52 33.87 & 41 21 48.90 & 5340 & 45 & 2 55 51.80 & 41 32 39.01 & 4224 & 77 \nl
2 52 38.50 & 41 34 41.10 & 4618 & 20 & 2 55 52.80 & 40 49 05.59 & 5727 & 31 \nl
2 52 42.30 & 41 19 33.50 & 4471 & 54 & 2 55 54.00 & 41 44 37.60 & 4317 & 94 \nl
2 52 43.10 & 41 06 08.76 & 5957 & 32 & 2 55 55.30 & 41 34 58.10 & 4748 & 27 \nl
2 52 51.95 & 41 26 21.30 & 5725 & 73 & 2 55 59.20 & 41 36 10.40 & 5957 & 67 \nl
2 52 58.90 & 41 32 01.60 & 6770 & 25 & 2 56 04.61 & 41 38 08.74 & 4466 & 29 \nl
2 52 59.00 & 41 58 24.71 & 5387 & 26 & 2 56 07.40 & 41 37 51.10 & 5948 & 26 \nl
2 53 05.61 & 41 04 27.94 & 4936 & 24 & 2 56 14.82 & 41 38 50.80 & 5234 & 59 \nl
2 53 07.70 & 41 25 03.68 & 5613 & 32 & 2 56 26.70 & 41 07 59.09 & 5092 & 41 \nl
2 53 16.20 & 41 29 11.80 & 5262 & 26 & 2 56 30.10 & 42 16 35.90 & 5882 & 53 \nl
2 53 20.75 & 41 56 48.30 & 5440 & 37 & 2 56 32.60 & 41 00 44.30 & 5596 & 19 \nl
2 53 26.28 & 41 56 35.20 & 5338 & 44 & 2 56 38.60 & 41 19 59.82 & 4633 & 10 \nl
2 53 30.00 & 41 39 03.38 & 4840 & 32 & 2 56 53.90 & 41 39 33.19 & 5374 & 45 \nl
2 53 40.20 & 41 43 31.10 & 6290 & 28 & 2 56 54.72 & 41 23 43.30 & 5594 & 60 \nl
2 53 44.10 & 41 27 15.08 & 4010 & 32 & 2 56 56.03 & 41 20 49.54 & 5153 & 41 \nl
2 53 47.10 & 41 32 49.09 & 5161 & 31 & 2 56 56.20 & 41 58 41.90 & 5331 & 17 \nl
2 53 50.00 & 41 27 18.40 & 4616 & 44 & 2 57 08.97 & 41 31 00.20 & 6334 & 31 \nl
2 53 51.30 & 41 43 38.20 & 5256 & 18 & 2 57 22.00 & 41 56 19.50 & 4439 & 20 \nl
2 53 54.50 & 41 40 25.61 & 5561 & 49 & 2 57 30.97 & 41 37 49.31 & 4498 & 42 \nl
2 53 59.50 & 41 47 35.09 & 6324 & 23 & 2 57 33.50 & 41 30 57.90 & 4943 & 21 \nl
2 54 03.46 & 41 33 29.00 & 6107 & 50 & 2 57 36.79 & 41 32 56.89 & 5121 & 23 \nl
2 54 05.72 & 41 57 44.90 & 4331 & 43 & 2 57 57.22 & 42 46 05.92 & 5338 & 35 \nl
2 54 14.80 & 41 23 00.00 & 6413 & 17 & 2 58 11.85 & 42 55 27.98 & 5417 & 44 \nl
2 54 16.40 & 41 39 38.41 & 5150 & 30 & 2 58 12.40 & 41 42 11.90 & 5658 & 26 \nl
2 54 16.98 & 41 34 52.30 & 5477 & 55 & 2 58 58.80 & 41 17 17.50 & 5065 & 25 \nl
2 54 24.40 & 41 36 19.40 & 6129 & 26 & 2 59 01.10 & 42 20 45.50 & 4954 & 22 \nl
2 54 25.30 & 41 34 35.90 & 5085 & 17 & 2 59 32.20 & 41 22 32.80 & 5753 & 20 \nl
2 54 26.85 & 41 39 19.33 & 5700 & 11 & 2 59 41.10 & 41 34 54.80 & 4956 & 29 \nl
2 54 27.40 & 41 30 47.30 & 4827 & 73 & 2 59 53.36 & 41 32 43.84 & 5625 & 23 \nl
2 54 27.50 & 41 34 42.50 & 5288 & 71 & 3 00 40.99 & 42 22 06.64 & 6013 & 25 \nl
2 54 28.56 & 41 35 58.00 & 5272 & 41 & 3 00 54.87 & 41 42 21.02 & 4911 & 32 \nl
2 54 28.57 & 41 26 55.94 & 4905 & 23 & 3 00 58.05 & 43 00 49.90 & 5025 & 25 \nl
2 54 29.04 & 41 10 07.60 & 4528 & 39 & 3 00 59.45 & 43 01 03.50 & 4828 & 22 \nl
2 54 30.40 & 41 36 36.22 & 6230 & 24 & 3 02 19.90 & 40 54 00.31 & 4703 & 98 \nl
2 54 34.01 & 41 33 31.27 & 5351 & 17 & 3 02 20.07 & 41 59 05.42 & 6736 & 25 \nl
2 54 36.00 & 41 53 25.00 & 5407 & 22 & 3 02 44.32 & 41 37 30.65 & 4976 & 30 \nl
2 54 38.30 & 41 35 18.10 & 5645 & 26 & 3 02 49.28 & 40 58 15.97 & 4851 & 54 \nl
2 54 39.25 & 42 09 36.93 & 6101 & 16 & 3 03 30.86 & 41 36 23.34 & 4797 & 28 \nl
2 54 40.00 & 41 34 31.12 & 5006 & 39 & 3 03 36.80 & 41 09 56.78 & 4871 & 21 \nl
2 54 41.40 & 41 33 49.28 & 5291 & 45 & \nodata & \nodata & \nodata & \nodata \nl
\enddata
\tablecomments{RA and Dec (J2000), and redshift and error (in \kms) for 
galaxies with measured redshifts.}
\label{tab:cz}
\end{deluxetable}


%

\begin{deluxetable}{lrrrrr}
\tablenum{2}
\tablewidth{0pc}
\tablecaption{Velocity Distribution}
\tablehead{
\colhead{Sample} &
\colhead{N$_{\rm gal}$} &
\colhead{$\overline{cz}$} &
\colhead{$\sigma_{\overline{cz}}$} &
\colhead{$\sigma_p$} &
\colhead{$\sigma_{200}$}     \\
\colhead{} & \colhead {} &
\colhead{[km s$^{-1}$]} & \colhead{[km s$^{-1}$]} & \colhead{[km s$^{-1}$]} &
\colhead{[km s$^{-1}$]} 
}
\startdata
Em+Ab      & 179 & 5186.2 &  54.3 &  $727.0^{+41.9}_{-35.8}$   & \nodata \\
Em         &  42 & 4821.5 & 168.0 & $1088.8^{+143.4}_{-103.0}$ & \nodata  \\
Ab$_{137}$ & 137 & 5247.6 &  49.0 &  $573.7^{+38.3}_{-32.0}$   & $601.5^{+46.1}_{-37.6}$ \\
Ab$_{135}$ & 135 & 5256.6 &  46.7 &  $542.9^{+36.6}_{-30.5}$   & $578.1^{+44.5}_{-36.3}$ \\


\tablecomments {Velocity dispersions are computed from galaxies within
1.72 \hmpc of the cluster center for Ab samples, and within 1.80 \hmpc
for Em and Em+Ab samples.  $\sigma_{200}$ is the projected velocity dispersion
within $r_{200}$ (\S 5.1).}
\enddata
\label{tab:velhist}
\end{deluxetable}



%

\begin{deluxetable}{lcccc}
\tablenum{3}
\tablecolumns{5}
\tablecaption{NFW $r_c$ from Projected Radii}
\tablehead{
 \colhead{Sample} &
 \colhead{$r_c$ / [ \hmpc ] } &
 \colhead{1$\sigma$ range} &
 \colhead{2$\sigma$ range} &
 \colhead{3$\sigma$ range}
}
\startdata
Ab$_{137}$     & 0.223 & 0.174 -- 0.285 & 0.157 -- 0.316 & 0.145 -- 0.342 \nl
Ab$_{135}$     & 0.211 & 0.164 -- 0.270 & 0.148 -- 0.299 & 0.137 -- 0.324 \nl
\tablecomments{$r_c$ is computed from two-zone magnitude-limited samples
of 123 and 121 members, respectively.}
\enddata
\label{tab:projrc}
\end{deluxetable}



%

\begin{deluxetable}{lccccccc}
\tablenum{4}
\tablecolumns{7}
\tablecaption{NFW Profile Parameters}
\tablehead{
 \colhead{Sample} &
 \colhead{$r_c$} &
 \colhead{$r_{200}$} &
 \colhead{$c$} &
 \colhead{$\delta_c$} &
 \colhead{$r_v$} &
 \colhead{M$_{200}$} &
 \colhead{M($<1.7$ \hmpc)}          \\
 \colhead{} &
 \colhead{[$h^{-1}$ Mpc]} &
 \colhead{[$h^{-1}$ Mpc]} &
 \colhead{} &
 \colhead{} &
 \colhead{[$h^{-1}$ Mpc]} &
 \colhead{[$10^{14}h^{-1}$M$_\odot$]}  &
 \colhead{[$10^{14}h^{-1}$M$_\odot$]} 
}
\startdata
Ab$_{137}$     & 0.223 & 1.02 & 4.6  & $7.1\times 10^3$ & 1.30 & 2.47 & 3.49 \nl
Ab$_{135}$     & 0.211 & 0.97 & 4.6  & $7.2\times 10^3$ & 1.28 & 2.13 & 3.01 \nl
\tablecomments{$r_c$ is computed from two-zone magnitude-limited samples
of 123 and 121 members, respectively.}
\enddata
\label{tab:nfwpar}
\end{deluxetable}



%

\begin{deluxetable}{lccc}
\tablenum{5}
\tablecolumns{4}
\tablecaption{NFW and Virial Masses}
\tablehead{
 \colhead{Sample} &
 \colhead{$M_{\rm NFW}(<1.7$ \hmpc)} &
 \colhead{$M_{\rm VT}(<1.7$ \hmpc)} &
 \colhead{$\Delta M / M_{\rm VT}$}                 \\
 \colhead{} &
 \colhead{[$10^{14}h^{-1}$M$_\odot$]} &
 \colhead{[$10^{14}h^{-1}$M$_\odot$]} &
 \colhead{} 
}
\startdata
Ab$_{137}$   & 3.49  &   3.10  &  0.126  \nl
Ab$_{135}$   & 3.01  &   2.74  &  0.099  \nl
\enddata
\label{tab:masscomp}
\end{deluxetable}


\clearpage

\clearpage

\clearpage

\clearpage

\clearpage

\clearpage

\clearpage

\end{document}